\documentclass[twocolumn]{autart}    

\usepackage{graphicx}          
\graphicspath{{Figure/}}
\usepackage{float}
\usepackage{caption}
\usepackage{subfig}
\usepackage{epstopdf}
\usepackage{amsmath}
\usepackage{multirow}
\usepackage{bm}
\usepackage{amsbsy}
\usepackage{amsfonts}
\usepackage{amsmath}
\usepackage{pdflscape}
\usepackage{amssymb}
\usepackage{enumitem}
\usepackage{color}
\usepackage{algorithm}
\usepackage{algorithmicx}

\makeatletter
\def\BState{\State\hskip-\ALG@thistlm}
\makeatother

\begin{document}

\begin{frontmatter}

\title{Robust Quadratic Gaussian Control of Continuous-time Nonlinear Systems}

\author[a]{Pouria Razzaghi}\ead{prazzaghi@smu.edu},    
\author[a]{Ehab Al Khtib}\ead{ealkhatib@smu.edu},               
\author[a]{Yildirim Hurmuzlu}\ead{Hurmuzlu@lyle.smu.edu}  

\address[a]{Dep. of Mechanical Engineering, Southern Methodist University, Dallas, TX, USA 75205}  


\begin{abstract}                          
In this paper, we propose a new Robust Nonlinear Quadratic Gaussian (RNQG) controller based on State-Dependent Riccati Equation (SDRE) scheme for continuous-time nonlinear systems. Existing controllers do not account for combined noise and disturbance acting on the system. The proposed controller is based on a Lyapunov function and a cost function includes states, inputs, outputs, disturbance, and the noise acting on the system. We express the RNQG control law in the form of a traditional Riccati equation. 

Real time applications of the controller places high computational burden on system implementation. This is mainly due to the nonlinear and complex form of the cost function. In order to solve this problem, this cost function is approximated by a weighted polynomial. The weights are found by using a least squares technique and a neural network. The approximate cost function is incorporated into the controller by employing a method based on Bellman's principle of optimality.

Finally, an inertially stabilized inverted pendulum example is used to verify the utility of proposed control approach. 
 
\end{abstract}

\end{frontmatter}

\section{Introduction}
Most systems are nonlinear in nature and subject to disturbances and noise. To control these systems, one needs a robust controller to deal with these undesired factors. In this paper, we focus on robust optimal controllers for nonlinear systems that are subject to disturbances, measurement and process noise. The main challenge of regulating this type of systems is accounting for noise and disturbance simultaneously in the control system design. In addition, on-line implementation of the resulting controllers can be too computationally intensive. For this purpose, we present a Neural Network (NN) approximation to render real time implementable controller.       


In a recent article, Gabriel \textit{et al.} \cite{gabriel2018optimal} presented a robust optimal controller for linear systems subject to disturbances. They, however, did not include the effect of noise in their analysis. Van Parys \textit{et al.} \cite{van2016distributionally} also investigated a robust control of constrained stochastic linear systems in the presence of disturbances. A development of robust optimal controller over a linear system with adjustable uncertainty sets was conducted in \cite{zhang2017robust}. Terra \textit{et al.} \cite{terra2014optimal} proposed an optimal robust recursive regulator for linear discrete-time systems that are subject to parametric uncertainties. In all these studies, underlying systems were linear and only the disturbance effect was considered.  

As far as nonlinear systems are concerned, several robust optimal control strategies were developed. Hu \textit{et al.} \cite{hu2016robust} presented a robust $H_{\infty}$ output-feedback control strategy for the path following of autonomous ground vehicles in the presence of parameter uncertainties and external disturbances. Yang and He \cite{yang2018self} proposed an Adaptive Dynamic Programming (ADP)-based self-learning robust optimal control scheme for input-affine continuous-time nonlinear systems with mismatched disturbances. They showed that to solve a Hamilton-Jacobi-Bellman-Isaacs (HJBI) equation, they needed a NN approximator. Satici \textit{et al.} \cite{satici2013robust} designed a robust $L_1$-optimal control of a quadrotor unmanned aerial vehicle (UAV).
Also, Zhang \textit{et al.} \cite{zhang2018event} studied a robust controller for uncertain nonlinear systems using ADP. They presented an event-based ADP algorithm by designing the NN weight updating laws. Wang \textit{et al.} \cite{wang2017h} presented a novel $H_2-H_{\infty}$ State-Dependent Riccati Equation (SDRE) control approach. They demonstrated the efficiency of control design framework for continuous-time nonlinear systems. All these studies deal with nonlinear underlying systems. They do consider effects of disturbances and uncertainties. But, the noise was not taken as a factor in their control system development. 

Also, this type of controller were implemented on both linear and nonlinear systems subjected to the noise. Bian and Jiang \cite{bian2016value} studied a robust optimal control on linear stochastic systems with input-dependent noise. Ma \textit{et al.} \cite{ma2015neural} investigated a class of nonlinear systems with external noises. They presented a NN-based adaptive robust controller to eliminate the effect of the external noises. Telen \textit{et al.} \cite{telen2015approximate} proposed an approximate robust optimal controller for nonlinear systems subjected to the process noise. They showed that their technique outperformed the unscented Kalman filter like techniques.

Most of the previous work considered systems subject to disturbances or noise separately. To the best of our knowledge, there is no previous study considered the combined effect of the disturbance and noise together in optimal robust control of a nonlinear system. Here, we propose a control law that can be applied to systems with noise and disturbance combined. Our proposed control law is based on nonlinear SDRE method. A general Lyapanov function based on system dynamics, inputs, outputs, disturbances, and noise is defined. Then, we derive a general Riccati equation to obtain a robust optimal feedback control law. Next, this  general Riccati equation is simplified to obtain an $H_2-H_{\infty}$ controller. Yet, implementing the conventional SDRE and our proposed controller in real-time applications requires high computational load \cite{cloutier1996nonlinear}. It becomes almost impossible for nonlinear systems subject to noise and disturbances. For this reason, we use the HJBI equation and NN weighted updating framework to define an approximate solution. In this way, it would be possible to use the proposed control scheme in real time. Our proposed control method can be used in different applications such as robotics, quadrotor UAVs,  mobile vehicles, and leader-follower systems. 

We apply the proposed controller to upright stabilization of a fly-wheeled inverted pendulum. This is a relevant example, because previous studies dealing with flywheel actuated inverted pendulum systems did not include the effect of the external disturbances and sensor noise \cite{hurmuzlu1998dynamics,borzova2004passively,hurmuzlu2004modeling,razzaghi2019nonlinear}. We consider the effect of both external disturbance and sensor noise on the system and demonstrate the efficiency of the proposed controllers.

This paper is organized in the following manner. First, In section \ref{controldesign}, we present the controller design. In section \ref{Dynamicmodel}, a mathematical model of the inverted pendulum is presented. Finally, section 4 presents the simulation results.

\section{Control Design} \label{controldesign}
Developing the SDRE based controller for a nonlinear system includes the following steps \cite{ccimen2008state}:

\begin{enumerate}
	\item  Define a cost function.
	\item  Express the dynamics in state-dependent coefficient form.
	\item  Solve the resulting Riccati equation.
	\item  Obtain the controller by using the traditional closed-loop feedback formula, with all the coefficients depending on the states instead of time.
\end{enumerate}

These steps for the control approach is presented in the next section. The procedure results in a Robust Nonlinear Quadratic Gaussian \textbf{(RNQG)} controller. This control law is based on a standard form of Riccati equation and does not increase the computational complexity of the problem. 
\subsection{Robust Nonlinear Quadratic Gaussian } \label{sec21}

In this paper, we focus on controlling the following nonlinear system:
\begin{align}
& \dot{\bar{x}}=f(\bar{x})+B(\bar{x})\,\bar{u}+F(\bar{x})\,\bar{w}+\bar{v} \label{system1} \\
& \bar{y}=C(\bar{x}) \, \bar{x}+ D(\bar{x})\, \bar{u}+ G(\bar{x})\,\bar{w}+ \bar{\epsilon} \label{system2}
\end{align}
where $\bar{x} \in\mathbb{R}^{n\times1}$, $\bar{u} \in \mathbb{R}^{m\times1}$, $\bar{y} \in \mathbb{R}^{r\times1}$, $\bar{w} \in \mathbb{R}^{q\times1}$,  $\bar{v} \in \mathbb{R}^{n\times1}$, and $ \bar{\epsilon}\in \mathbb{R}^{r\times1} $ are the state of the system, the input applied to system, the output, the external disturbance acting on the system, the process noise, and the measurement noise vectors, respectively. The notation $R^{n\times 1}$ indicates that the matrix $R$ is a $n\times 1$ matrix. 

Based on extended linearization \cite{friedland1995advanced}, and under the assumption $f(\bar{0})=\bar{0}$, a continuous nonlinear matrix-valued function $A(\bar{x})$ always exists such that:
\begin{equation}
f(\bar{x})= A(\bar{x})\,\bar{x}
\end{equation}
So we can rewrite the system dynamics as  follows: 
\begin{align}
& \dot{\bar{x}}=A(\bar{x})\,\bar{x}+B(\bar{x})\,\bar{u}+F(\bar{x})\,\bar{w}+\bar{v} \label{system3} \\
& \bar{y}=C(\bar{x}) \, \bar{x}+ D(\bar{x})\, \bar{u}+ G(\bar{x})\,\bar{w}+ \bar{\epsilon} \label{system4}
\end{align}
where $A^{n\times n},B^{n\times m},C^{r\times n},D^{r\times m},F^{n\times q}$, and $G^{r\times q}$ are the known state dependent coefficient matrices. We should note that in Eq.~(\ref{system3}), the vector $\bar{v}$ is a stochastic process called process noise. Its mathematical characterization is:
\begin{equation}
E(\bar{v})=0
\end{equation}
where the function $E(.)$ is the expected value \cite{sinha2007linear}, and:
\begin{equation}
E(\bar{v}(t)\bar{v}^T(t+\tau))=L v(\tau)
\end{equation}
where the matrix $L^{n\times 1}$ is called the intensity matrix of the process noise with the property $ L\,L^T >0 $ and $v(\tau)$ is a Dirac delta function \cite{sinha2007linear}. Also, in Eq.~(\ref{system4}), the vector $\bar{\epsilon}$ is a measurement noise, which can be the noises from sensor's extracted data. It is assumed that: 
\begin{equation}
E(\bar{\epsilon}(t))=0 ; \,\,\,\,  E(\bar{\epsilon}(t)\bar{\epsilon}^T(t+\tau))=H v(\tau)
\end{equation}
where the matrix $H^{r\times 1}$ is called the intensity matrix of the measurement noise with the property $ H,H^T > 0 $. It is also assumed that process and measurement noise vectors are uncorrelated \cite{sinha2007linear};
\begin{equation}
E(\bar{v}(t)\bar{\epsilon}^T(t+\tau))=E(\bar{\epsilon}(t)\bar{v}^T(t+\tau))=0
\end{equation} 
Consider a Lyapunov function given by:
\begin{equation}
V= \bar{x}^T P(\bar{x})\, \bar{x} \geq 0
\end{equation}
where $P(\bar{x}) ^{n \times n}$ is a positive definite matrix. The corresponding cost function $J$ for this case can be written as:
\begin{align}
J=V(t_f)&+ \int_{0}^{t_f}\left(\bar{x}^TQ(\bar{x})\, \bar{x}+\bar{u}^TR(\bar{x})\,\bar{u}+ \bar{y}^TS(\bar{x})\bar{y}\right) dt\nonumber\\
&+ \gamma_1^2\int_{0}^{t_f} \bar{w}^T\bar{w} \, dt+\gamma_2^2\int_{0}^{t_f} v^Tv \, dt
\end{align}
where $Q^{n \times n}$, $R^{m \times m}$, and $S^{r \times r}$ are symmetric positive definite matrices and $\gamma_1 , \gamma_2$ are real numbers. The derivative of cost function can be written as: 
\begin{align}
\dot{J}&=\dot{V}+ \bar{x}^TQ(\bar{x})\, \bar{x}+\bar{u}^TR(\bar{x})\,\bar{u}+ \bar{y}^TS(\bar{x})\bar{y} \nonumber\\
&+ \gamma_1^2\, \bar{w}^T\bar{w}+\gamma_2^2\, v^Tv\leq 0 \label{perf}
\end{align} 
Our goal is to provide optimality conditions for the determination of a set of state feedback gains $K^{m \times n}$ such that:
\begin{equation}
\bar{u}= K(\bar{x})\, \bar{x}
\end{equation} 
and the closed-loop system is stable and a desired cost function (Eq.~(\ref{perf})) is minimized. To find the optimal gains ($K_o$), we need to identify the components of the matrix $P$ in the Lyapunov function. By substituting the Lyapunov function, the system model, output, and control input equations in $\dot{J}$, Eq.~(\ref{perf}) yields:
\begin{align}
\bar{x}^T P\, \dot{\bar{x}}&+\dot{\bar{x}}^TP\, \bar{x} + \bar{x}^T\dot{P}\, \bar{x}+ \bar{x}^TQ\, \bar{x}+\bar{u}^TR\,\bar{u}+ \bar{y}^TS\,\bar{y}\nonumber \\
&+\gamma_1^2\, \bar{w}^T\bar{w}+\gamma_2^2\, v^Tv\leq 0 \\
\Rightarrow  \bar{x}^T P\,&[A\,\bar{x}+B\,\bar{u}+F\,\bar{w}+L\, v]\nonumber \\
&+ [A\,\bar{x}+B\,\bar{u}+F\,\bar{w}+L\, v]^T\, P\, \bar{x}\nonumber \\
& +\bar{x}^T \dot{P}\, \bar{x}+ \bar{x}^TQ\,\bar{x}+\bar{x}^TK^TR\,K\, \bar{x} \nonumber \\
& + [C \, \bar{x}+ D\, K\, \bar{x}+ G\,\bar{w}+ H\,v]^T S \nonumber \\
& [C \, \bar{x}+ D\, K\, \bar{x}+ G\,\bar{w}+ H\,v] \nonumber \\
&+\gamma_1^2\, \bar{w}^T\bar{w}+\gamma_2^2\, v^Tv\leq 0 \label{eq14}
\end{align}
Equation (\ref{eq14}) can be equivalently rewritten in matrix form as:
\begin{align} \label{eq1}
\bar{\xi}^T \boldsymbol{M}& \bar{\xi} \leq 0 \rightarrow 
\begin{bmatrix}
\bar{x} &\bar{w}  &v 
\end{bmatrix}\begin{bmatrix}
M_{1} &M_2  &M_3 \\ 
M_2^T& M_4 & M_5\\ 
M_3^T&M_5^T  &M_6 
\end{bmatrix}\begin{bmatrix}
\bar{x}\\ 
\bar{w}\\ 
v
\end{bmatrix} \leq 0 \\
M_1&=P\,[A+B\,K] + [A+B\,K]^T\, P+ \dot{P}+ Q \nonumber \\
&+K^T\,R\,K +[C + D\, K]^TS\,[C+ D\, K]\nonumber \\
M_2&=P\,F+[C + D\, K]^T S\, G \nonumber \\
M_3&= P\,L +[C + D\, K]^T S\, H \nonumber \\
M_4&= G^TS\,G+\gamma_1^2\, I_6;\:\: M_5= G^TS\,H \nonumber \\
M_6&= H^TS\,H + \gamma_2^2\,I_1 \nonumber
\end{align}
where $I_{i}$ is the $i$-dimension identity matrix. We should note that the dimensions of the matrices $\boldsymbol{M}, M_1, M_2, M_3, M_4, M_5$ and $M_6$ are $(n+q+1)\times (n+q+1)$, $n\times n$, $n \times q$, $n \times 1$, $q \times q$, $q \times 1 $, and $1\times 1$, respectively. 

\textbf{Definition.} In linear algebra, the Schur complement \cite{zhang2006schur} of a symmetric matrix ($N$) is defined as: 
\begin{equation}
N=\begin{bmatrix}
N_1 &N_2 \\ 
N_2^T &N_3  
\end{bmatrix}
\end{equation} 
where $N_1$, $N_2$, and $N_3$ are respectively $ n \times n $, $  n \times m $, and $ m \times m $ matrices, and $N_3$ is invertible.

In addition, the Schur complement of the block $N_3$ of the matrix $N$ is the $ n \times n $ matrix defined by:
\begin{equation}
N/N_3:= N_1 -N_2 \,N_3^{-1}N_2^T
\end{equation} 
Then, $N$ is negative definite if and only if $N$ and $N/N_3$ are both negative definite. 
We use this definition for the matrix $\boldsymbol{M}$ and the matrix blocks $N_1$, $N_2$, and $N_3$ are extracted from Eq.~(\ref{eq1}) as:
\begin{align}
& N_1=\begin{bmatrix}
M_1 & M_2 \\ 
M_2^T&M_4 
\end{bmatrix}  ; \:\:
N_2= \begin{bmatrix}
M_3\\ 
M_5
\end{bmatrix}; \:\: N_3= M_6
\end{align} 
Thus the Schur complement $Z=N/N_3$ becomes:
\begin{align}
& Z= N_1 -N_2\, N_3^{-1}N_2^T= \begin{bmatrix}
Z_1 &Z_2 \\ 
Z_2^T& Z_3
\end{bmatrix} \\
& Z_1= M_1 -M_3\, M_6^{-1}M_3^T \nonumber \\
& Z_2= M_2 - M_3 \, M_6^{-1}M_5^T \nonumber \\
& Z_3= M_4- M_5\, M_6^{-1}M_5^T
\end{align}
noting that the matrix $Z^{(n+q)\times(n+q)}$ is symmetric. By applying the Schur complement results on matrix $Z$ again, we have the following $n \times n$ matrix inequality:
\begin{align}
&  M_1 -M_3\, M_6^{-1}M_3^T-[M_2 - M_3 \, M_6^{-1}M_5^T ] \nonumber \\
&  [M_4- M_5\, M_6^{-1}M_5^T]^{-1}[M_2 - M_3 \, M_6^{-1}M_5^T ]^T \leq 0 \label{25}
\end{align} 
By substituting Eq.~(\ref{eq1}) in to (\ref{25}), we equivalently have:
\footnotesize{
	\begin{flalign}
	[P\,&[A+B\,K] + [A+B\,K]^T\, P+Q+K^T\,R\,K+\nonumber \\
	&[C + D\, K]^TS\,[C+ D\, K]]- \nonumber \\
	& [[P\,L +[C + D\, K]^TS\, H][H^TS\,H+ \gamma_2^2\,I_1 ]^{-1}\nonumber \\
	& [P\,L +[C + D\, K]^TS\, H]^T]- [[P\,F+[C + D\, K]^TS\, G]-\nonumber \\
	& [P\,L +[C + D\, K]^TS\, H][H^TS\,H+ \gamma_2^2\,I_1  ]^{-1}H^TS\,G ]\nonumber \\
	& [[G^TS\,G+\gamma_1^2\, I_6]-G^TS\,H[H^TS\,H+ \gamma_2^2\,I_1  ]^{-1}H^TS\,G]^{-1}\nonumber \\
	& [[P\,F+[C + D\, K]^TS\, G]-[P\,L +[C + D\, K]^TS\, H] \nonumber \\
	&  [H^TS\,H + \gamma_2^2\,I_1 ]^{-1}H^TS\,G]^T\leq -\dot{P} \label{eq2}
	\end{flalign}}\normalsize 
In order to guarantee stability, the Lyapunov function must be decreasing ($\dot{V} \le 0$), which results in $ 0 \leq -\dot{P}$. Thus, the left hand side of the inequality Eq.~(\ref{eq2}) becomes equal to zero. By grouping the different terms of $ K $ in Eq.~(\ref{eq2}), we have:\footnotesize{
	\begin{align}
	\Gamma_1+& \Gamma_2 K + K^T \Gamma_2^T + K^T \Gamma_3 K=0 \label{eq3}\\
	\Gamma_1=& P\,A +A^TP + Q+C^TS\,C- \nonumber\\
	&(P\,L+C^TS\,H)M_6^{-1}(P\,L+C^TS\,H)^T-\nonumber \\
	& [(P\,L+C^TS\,H)M_6^{-1}H^TS\,G-(PF+C^TS\,G)]\Gamma_4 \nonumber\\
	& [(P\,L+C^TS\,H)^TM_6^{-1}H^TS\,G-(PF+C^TS\,G)]^T \nonumber \\ 
	\Gamma_2=& P\,B + C^TS\, D -(P\,L+C^TS\,H)M_6^{-1} H^T S\,D-\nonumber \\
	&[(P\,L+C^TS\,H)M_6^{-1}H^TS\,G-(PF+C^TS\,G)]\nonumber\\
	&\Gamma_4 \,G^T[I_r-S\,H\,M_6^{-1}H^T]S\,D \nonumber \\
	\Gamma_3= &R+ D^TS\,D- D^TS\,H\,M_6^{-1}H^TS\,D-D^TS[G\,\Gamma_4\,G^T-\nonumber \\
	&G\,\Gamma_4\,G^TS\,H\,M_6^{-1}H^T -H\,M_6^{-1}H^TS\,G\,\Gamma_4\,G^TS+\nonumber \\
	& H\,M_6^{-1}H^TS\,G\,\Gamma_4\,G^TS\,H\,M_6^{-1}H^T]S\,D \nonumber \\
	\Gamma_4=&[[G^TS\,G+\gamma_1^2\, I_6]-G^TS\,H[H^TS\,H+ \gamma_2^2\,I_1  ]^{-1}H^TS\,G]^{-1}\nonumber
	\end{align}}\normalsize 
By completing the square \cite{iwasaki1995parametrization} in gain $ K $ and use the optimal gain $K_o$, we have:
\begin{equation}
\Gamma_1+(K-K_o)^T \Gamma_3(K-K_o)-K_o^T \Gamma_3 K_o=0 \label{eq4}
\end{equation} 
We should note that the matrices $\Gamma_3$ and $ \Gamma_3^{-1} $ are symmetric, so by comparing Eqs.~(\ref{eq3}) and (\ref{eq4}), the optimal feedback gain should be:
\begin{equation}
K_o= -\Gamma_3^{-1}\Gamma_2^T  \label{optimalgain}
\end{equation} 

We substitute $K=K_o$ in Eq.~(\ref{eq4}), then the matrix $P$ can be calculated by the solution of the following equation:\footnotesize{
	\begin{align}
	\Gamma_1 -& K_o^T\Gamma_3K_o = \Gamma_1 - \Gamma_2\Gamma_3^{-1}\Gamma_2^T =P\,A +A^TP + Q+ \nonumber\\
	&C^TS\,C-(P\,L+C^TS\,H)M_6^{-1}(P\,L+C^TS\,H)^T-\nonumber \\
	& [(P\,L+C^TS\,H)M_6^{-1}H^TS\,G-(PF+C^TS\,G)]\Gamma_4 \nonumber\\
	& [(P\,L+C^TS\,H)^TM_6^{-1}H^TS\,G-(PF+C^TS\,G)]^T - \nonumber\\
	&(P\,B + C^TS\, D -(P\,L+C^TS\,H)M_6^{-1} H^T S\,D-\nonumber \\
	& [(P\,L+C^TS\,H)M_6^{-1}H^TS\,G-(PF+C^TS\,G)]\nonumber\\
	&\Gamma_4 \,G^T[I_r-S\,H\,M_6^{-1}H^T]S\,D)(R+ D^TS\,D-\nonumber\\
	& D^TS\,H\,M_6^{-1}H^TS\,D-D^TS[G\,\Gamma_4\,G^T-\nonumber \\
	& G\,\Gamma_4\,G^TS\,H\,M_6^{-1}H^T -H\,M_6^{-1}H^TS\,G\,\Gamma_4\,G^TS+\nonumber \\
	& H\,M_6^{-1}H^TS\,G\,\Gamma_4\,G^TS\,H\,M_6^{-1}H^T]S\,D)^{-1}\nonumber \\
	&(P\,B + C^TS\, D -(P\,L+C^TS\,H)M_6^{-1} H^T S\,D-\nonumber \\
	& [(P\,L+C^TS\,H)M_6^{-1}H^TS\,G-(PF+C^TS\,G)]\nonumber\\
	&\Gamma_4 \,G^T[I_r-S\,H\,M_6^{-1}H^T]S\,D)^T=0
	\end{align} }\normalsize
Although this generalized Riccati equation seems quite complicated, we need to rewrite the entire equation in the form of conventional Riccati equation. This algebraic transformation helps to use the standard Riccati solver and find the optimal gain from Eq.~(\ref{optimalgain}). Following notations are defined to obtain the standard Riccati equation.\footnotesize{ 
\begin{flalign}
& \lambda_1=(\Gamma_4 \,G^T[I_r-S\,H\,M_6^{-1}H^T]S\,D) \nonumber \\
& \lambda_2= \Gamma_3^{-1} \nonumber \\
& \lambda_3= [L\,M_6^{-1}H^TS\,G-F] \nonumber \\
& \lambda_4= [C^TS\,H\,M_6^{-1}H^TS\,G-C^TS\,G] \nonumber \\
& \lambda_5= C^TS\, D -C^TS\,H\,M_6^{-1} H^T D \nonumber \\
& \lambda_6=B- L\,M_6^{-1} H^TS\, D \nonumber \\
& \Lambda_1= -L\,M_6^{-1}H^TS\,C+ \lambda_3\, \Gamma_4 \,\lambda_4^T- \nonumber \\
&\: \: \: \: \: \: \: \: \: \:[\lambda_6- \lambda_3\,\lambda_1]\, \lambda_2 [\lambda_5- \lambda_4\, \lambda_1]^T \nonumber \\
& \Lambda_2=Q+C^TS\,C-C^TS\,H\,M_6^{-1}H^TS\,C+ \lambda_4\,\Gamma_4 \lambda_4^T - \nonumber \\
&\: \: \: \: \: \: \: \: \: \:  [\lambda_5- \lambda_4\,\lambda_1]\,\lambda_2  [\lambda_5- \lambda_4\,\lambda_1]^T \nonumber \\
& \Lambda_3=L\,M_6^{-1}L^T+ \lambda_3\,\Gamma_4  \lambda_3^T -  [\lambda_6 -\lambda_3\,\lambda_1]\,\lambda_2  [\lambda_6 -\lambda_3\lambda_1]^T 
\end{flalign} }\normalsize
Then the conventional form of generalized Riccati equation can be presented as follows:
\begin{flalign}
P(A+\Lambda_1)+ (A+\Lambda_1)^TP+ \Lambda_2 + P \, \Lambda_3\, P^T=0
\end{flalign}

If we do not have the noise on the system, that is, $L=H=0$, and only $H_{\infty}$ performance criterion exists, then the general controller can be converted to $H_2-H_{\infty}$ control \cite{wang2017h}. The optimal feedback gain can be found as follows:
\begin{flalign}
E_1= &[G^TS\,G+\gamma_1^2\, I_6]^{-1} \nonumber \\
K_o =& -[R- D^TS\,G\,E_1\,G^TD +D^TS\,D]^{-1}\nonumber \\
& [P\,B + C^TS\, D -(PF+C^TS\,G)E_1 \,G^TS\,D ]^T
\end{flalign}

In addition, the solution of conventional SDRE control can be regulate as follows. The simplest system model and its cost function used in the solution of the benchmark problem is:
\begin{flalign}
&  \dot{\bar{x}}=A(\bar{x})\,\bar{x}+B(\bar{x})\,\bar{u} \\
&J=\int_{0}^{t_f} [\bar{x}^T Q(\bar{x})\, \bar{x} + \bar{u}^T R(\bar{x})\, \bar{u}] dt \label{Cost}
\end{flalign} 
The solution of the optimal control to minimize the cost function is obtained by solving the following Riccati equation: 
\begin{flalign} \label{SDRERiccati}
A^T(\bar{x})\,& P(\bar{x}) + P(\bar{x})\,A(\bar{x})-\nonumber \\
&P(\bar{x})\, B(\bar{x})\, R^{-1}(\bar{x})\, B^T(\bar{x})\, P(\bar{x}) + Q(\bar{x})=0    
\end{flalign}

The resulting Riccati equation solution is a function of the states. The optimal feedback controller is:
\begin{equation} \label{uSDRE}
\bar{u}=-R^{-1}(\bar{x})\, B^T(\bar{x})\, P(\bar{x})\,\bar{x}    
\end{equation} 
%
The SDRE controller, by its construction, ensures that there is a near optimal solution for the system \cite{ccimen2010systematic}. 

In implementing SDRE approach, the most desirable option is to solve Eq.~(\ref{SDRERiccati}) in a closed form. This may be possible for some systems having special forms. In general, however, an analytical solution cannot be obtained. In that case, the second option is to obtain numerical solution of the problem in real time. The time increment of the discretized solutions of Eqs.~(\ref{SDRERiccati}) and (\ref{uSDRE}) can be automatically set by a simple Euler or the Runge-Kutta routine.

In general, real time implementation of the SDRE is computationally taxing \cite{cloutier2002capabilities}. This is particularity true for high order systems. Thus, we propose an approximation method that can highly speed up the real time computational, which is presented in next subsection.

\subsection{The Approximation Method}

We seek an approximation $ \widetilde{u}(\bar{x}) $ to the SDRE controller such that it is also a solution of the  HJBI equation \cite{feng2009game}. For this purpose, a positive definite cost function $ V(\bar{x}) $ is required such that it satisfies the following equation:
\begin{equation}
\underset{u}{min}\left \{{u}^T \frac{R(\bar{x})}{2}u+\triangledown V^T(\bar{x})\dot{\bar{x}}\right\}+\widetilde{Q}(\bar{x})=0 
\label{eq31} 
\end{equation}
where $\triangledown V$ is the partial derivative of function $V$ with respect to $\bar{x} $. We should note that $\widetilde{Q}(\bar{x}) $ is a positive definite, which may be differ from $ Q(\bar{x})$ in Eq.~(\ref{Cost}).
Here, we consider two sets of equations of motion for SDRE and RNQG control systems given by:
\begin{flalign}
& \dot{\bar{x}}= f(\bar{x})+ B(\bar{x}) u \label{SDREEq}\\
& \dot{\bar{x}}= f(\bar{x})+ B(\bar{x}) u+ F(\bar{x})\,\bar{w}+\bar{v} \label{RNQGEq}
\end{flalign}
The minimization in Eq.~(\ref{eq31}) can be readily obtained by:
\begin{equation}
\widetilde{u}(\bar{x})=-R^{-1}(\bar{x})g(\bar{x}) \triangledown V(\bar{x}) \label{uapx}
\end{equation}
which when substituted into the HJBI equation to yield:
\begin{equation}
\widetilde{Q}(\bar{x})- \triangledown{V}^T(\bar{x})g(\bar{x})\frac{R^{-1}(\bar{x})}{2}{g}^T(\bar{x})\triangledown V+ \triangledown{V}^T(\bar{x})f(\bar{x})=0 \label{approx}
\end{equation}

We should note that, one can find the approximation of SDRE and RNQG by substituting $\dot{\bar{x}}$ with coressponding terms in Eqs.~(\ref{SDREEq}) and (\ref{RNQGEq}), respectively. The function $ \widetilde{Q}(\bar{x}) $ is considered as an unknown and the function $ R(\bar{x}) $ is taken as known. In general, there does not exist a function $ V(\bar{x}) $ such that SDRE controller (Eq.~(\ref{uSDRE})) is expressible as Eq.~(\ref{uapx}). Ideally the approximation problem should be combined with the calculation of $ V(\bar{x}) $ and $ \widetilde{Q}(\bar{x}) $ in a single step in Eq.~(\ref{approx}). This however leads to a difficult problem. A necessary condition that $ \widetilde{u}(\bar{x}) $ should satisfy is that it should be expressible in the form of Eq.~(\ref{uapx}). In this paper, the approach is to find a polynomial function $ V_{0}(\bar{x}) $, which when substituted for $ V(\bar{x}) $ in Eq.~(\ref{uapx}), the result gives a controller $ \widetilde{u}(\bar{x}) $, which is the approximation of controllers in the least square sense. The approximation in the least squares sense is chosen in the interest of simplicity and tractability. The coefficients in the polynomial function $ V_{0}(\bar{x}) $ is determined by following a learning process, which will be explained next. To explain the idea of approximation method, the system in Eq.~(\ref{SDREEq}) is represented in the discrete form and an algorithm is proposed. Let consider the dynamics of a system in the discrete form as:
\begin{equation}
\bar{x}_{k+1}=f(\bar{x}_{k})+ g(\bar{x}_{k}) u_k
\label{eq17}
\end{equation}
where $k$ is the discrete time variables. The cost function in the discrete time form and its recurrence equations can be written as:
\begin{flalign}
& J=\Theta (\bar{x}_{N})+\sum_{k=0}^{N-1}\Theta(\bar{x}_{k}) \label{eq18} \\
& J= J_{N}(\bar{x}_{N})+ \sum_{k=0}^{N-1} J_{k}(\bar{x}_{k})
\end{flalign}
where
\begin{flalign}
& J_{N}(\bar{x}_{N})=\Theta (\bar{x}_{N}), \\
& J_{k}(\bar{x}_{k})=\Theta(\bar{x}_{k})+J_{k+1}(\bar{x}_{k+1}) 
\end{flalign}
where $N$ denotes the final time. The objective is to approximate the function $J _{k} (\bar{x}_{k})$ as a polynomial of the states. A neural network ($NN$) as an approximator is trained for this purpose. Moreover, with the function approximation, the cost function for the network can be written as: 
\begin{equation} 
J_{k}(\bar{x}_{k})\cong W_{k}^{T} \Upsilon  (\bar{x}_{k})
\end{equation}
where $W_{k}$ is the unknown optimal weights of the network at time step and $ \Upsilon$ is the basis function of states. The training process for weights $ W_{k} $, is presented in Algorithm~\ref{ALg1}. Once the algorithm converges, the cost function is approximated by $W_k^T\,\Upsilon(\bar{x}_k)$ in a closed form.

\begin{algorithm}
	\caption{}\label{ALg1}
	\begin{algorithmic}[1]
		\State $k \gets N$
		\State $W_{N+1}^T = \bar{0}$
		\BState \emph{loop}:
		\State Select random $\eta$ different states, $ \bar{x}_{k}^{(j)}$, $ j\in \{1, 2,\cdots ,\eta \}$ for $ \eta \gg 1 $ 
		
		\State Train network weights for $W_k$ such that
		\begin{equation}\label{eq41}
		W_k^T\,  \Upsilon( \bar{x}_k^{(j)})=\Theta  (\bar{x}_{k}^{(j)})+W_{k+1}^T\, \Upsilon(f( \bar{x}_k^{(j)}))
		\end{equation}
		\State $k \gets k-1$
		\State \textbf{go to} \emph{loop}.
	\end{algorithmic}
\end{algorithm}

One may use the method of least squares to find the unknown weights as coefficients of the selected polynomial in the training process \cite{heydari2014optimal}. For this purpose, $ \eta $ random states should be selected to apply in the least squares method. Using Eq.~(\ref{eq41}) yields:  
\begin{equation}
\left\{\begin{matrix}
W^{T}\Upsilon (x^{[1]})=\nu ( x^{[1]})\\ 
\vdots \\ 
W^{T}\Upsilon (x^{[\eta ]})=\nu ( x^{[\eta ]})
\end{matrix}\right.\label{eq42}
\end{equation}
where $\nu ( x^{[j]}) = \Theta  (\bar{x}^{(j)})+W^T\, \Upsilon(f( \bar{x}^{(j)}))$. If we define $ \boldsymbol{\Upsilon} \equiv [\Upsilon (x^{[1]}), \Upsilon (x^{[2]}), \cdots, \Upsilon (x^{[\eta ]})]$ and $ \boldsymbol{\nu}\equiv [
\nu (x^{[1]}), \nu (x^{[2]}), \cdots,\nu (x^{[\eta ]})] $ and using the method of least squares, the solution of Eq.~(\ref{eq42}) is:
\begin{equation}
\boldsymbol{W}=(\boldsymbol{\Upsilon}\boldsymbol{\Upsilon}^{T})^{-1}\boldsymbol{\Upsilon} \boldsymbol{\nu} ^{T}
\end{equation}

Once the neural network controller trained through Algorithm~\ref{ALg1}, it can be utilized for real time optimal control of the system. The controller can be implemented in real time by substituting $\bar{x}_k$ at each time step into Eq.~(\ref{eq31}) to calculate $\widetilde{u}$ and applying it to the system. The convergence and the stability of Algorithm~\ref{ALg1} can be found in \cite{heydari2014global}.

\section{application to a nonlinear system} \label{Dynamicmodel}
\subsection{System Description}

\begin{figure}[h]
\begin{center}
	\includegraphics[width=6cm]{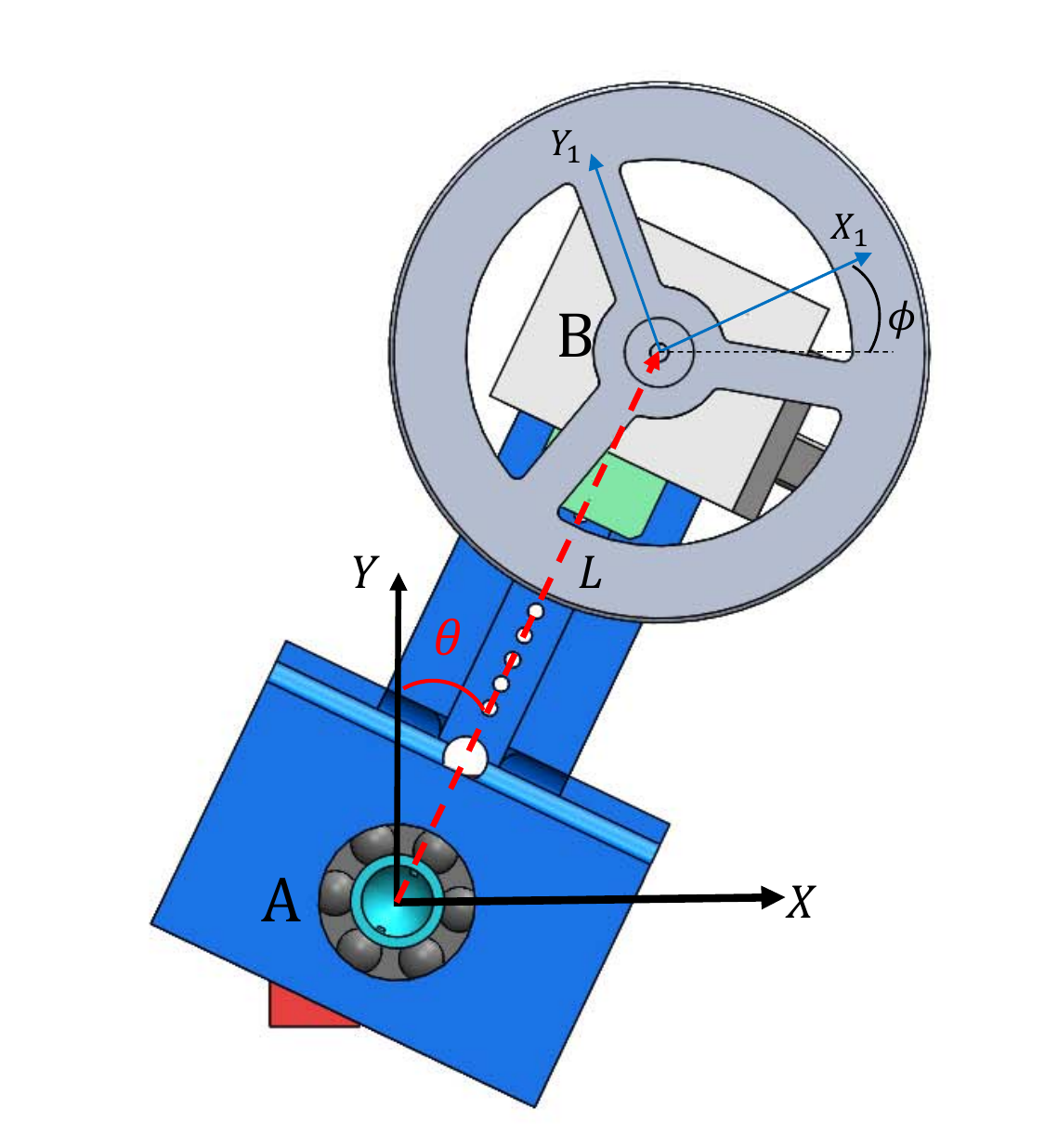} 
	\caption{Schematics of the inverted pendulum with the reference frames.}
	\label{Results}
\end{center}
\end{figure}

The inverted pendulum considered here is presented in Fig.~\ref{Results}. As shown, the pivot point of the pendulum is at point A. The joint at A is free to rotate and is not actuated. A flywheel is attached to the tip of the pendulum. The flywheel is driven by a brushless DC motor attached at point B. The control objective is to stabilize the inverted pendulum at its vertical position by using inertial actuation generated by the flywheel. Typically IMU sensors are used to measure the angular position ($\theta$) of the pendulum. Such sensors are often prone to measurement noise.   

The mass of the pendulum is $M_{p}$ and $ M_w $ is the pendulum mass. The parameters $L_e$ and $L_{G}$ are the lengths of the elbow and the distance between the rotation center and the center of gravity of the pendulum, respectively. Figure~\ref{Results} depicts the reference frame \textit{XY}, and the body frame $X_{1}Y_{1}$, placed at the centers of rotation and the flywheel, respectively. The state variables of the system are $\{\theta,\,\phi\}$, where $\theta$ is pendulum angle and $\phi$  is the rotation angle of the flywheel in the counter-clockwise direction.

\subsection{Equations of the motion} 
The equations of motion are derived using Lagrange's method~(\ref{lagrang}). The equations of motion can be obtained by substituting the total kinetic ($T_{total}$) and potential ($V_{total}$) energies of the system into the Lagrangian ($\mathfrak{L}$) equation.
\begin{eqnarray}
&&\mathfrak{L} = T_{total} - V_{total}; \,\,\frac{\partial }{\partial t } [\frac{\partial \mathfrak{L}}{\partial \dot{q_{i}}}] - \frac{\partial \mathfrak{L} }{\partial q_{i}} = Q_{i}	
\label{lagrang}
\end{eqnarray}
where $q_i$ and $Q_{i}$ are the $i^{th}$  generalized coordinate and generalized force, respectively. Total kinetic and potential energy expressions of the system can be written as:
\small{\begin{flalign}
	&T_{total}=\frac{1}{2} (M_p L_G^2+M_w L_e^2 + I_p+I_w) \dot{\theta}^2+ I_w \dot{\theta} \dot{\phi}+ \frac{1}{2} I_w \dot{\phi}^2
	\label{Kinetic} \\
		&V_{total}=(M_p L_G + M_w L_e) g \cos \theta
	\label{Potential}
	\end{flalign}
}\normalsize
where $I_p$ is the moment of inertia of the pendulum and $I_w$ is the moment of the inertia of flywheel. The equations of the motion of the system are given by:
\begin{flalign}
	(M_p L_G^2+&M_w L_e^2 + I_p+I_w)\ddot{\theta}+ I_w \ddot{\phi} \nonumber \\
	& -(M_p L_G + M_w L_e) g\sin \theta=0 \label{EQMotion1} \\
	I_w(\ddot{\theta}+&\ddot{\phi})= T_w \label{EQMotion2} 
	\end{flalign}
where $T_w$ is the flywheel drive torque as the input to the system. In the next step, to control the flywheel at the desired speed, the mathematical model of the motor's physical behavior is included in the system as follows:
\begin{equation}
V= L_m \frac{di}{dt} + R_m i + K_e \omega_m; \,\,\, T_w= N_g K_t i  \label{motor2} 
\end{equation}
where $V$ is the motor voltage and $i$ is the armature current. $L_m $ and $R_m$ are the armature coil inductance and resistance, respectively. The motor back electro-magnetic force is given bu $K_e$ and $\omega_m$ is the angular velocity of the motor. The gear ratio and the motor torque constant are given by $N_g$ and $K_t$, respectively. Using the relationship between the motor and the flywheel, we can calculate the required motor voltage in terms of the flywheel angular velocity. 

The state-space vector is defined as $\bar{x}=\{\theta,\,\phi,\,\dot{\theta},\,\dot{\phi}\}$ and the equations of motion is represented in the state-space form as:
\begin{flalign}
& \dot{\bar{x}}= \bar{F}( \bar{x}  )+ \bar{G}\, \textit{u}  \label{SSEQ} \\
&\begin{bmatrix}
\dot{x}_1\\ 
\dot{x}_2\\ 
\dot{x}_3\\ 
\dot{x}_4 
\end{bmatrix}=\begin{bmatrix}
x_3\\ 
x_4\\ 
\frac{C_T}{I_t} \sin x_1  \\ 
-\frac{C_T}{I_t} \sin x_1 
\end{bmatrix} + \begin{bmatrix}
0\\ 
0\\ 
-\frac{1}{I_T}\\ 
\frac{(I_T+I_w)}{I_w I_T}
\end{bmatrix} T_w \label{SSEQ1} \\
& C_T= (M_p L_G + M_w L_e) g; \,\, I_T= M_p L_G^2+M_w L_e^2 + I_p \nonumber
\end{flalign}

\section{Simulation results} \label{experiment}
\begin{figure}[h!]
\begin{center}
	\subfloat[The pendulum angle.]{{\label{Simtheta1}}{\includegraphics[width=8cm]{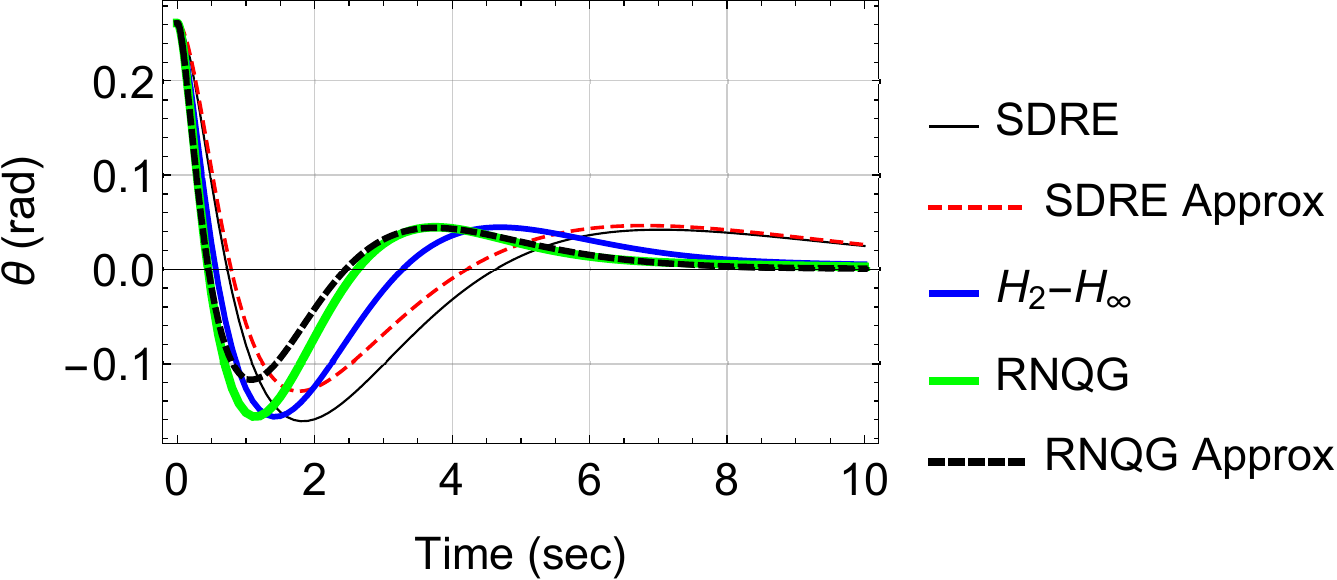} }}%
	\qquad
	\subfloat[The pendulum angle rate.]{{\label{Simdtheta1}}{\includegraphics[width=8cm]{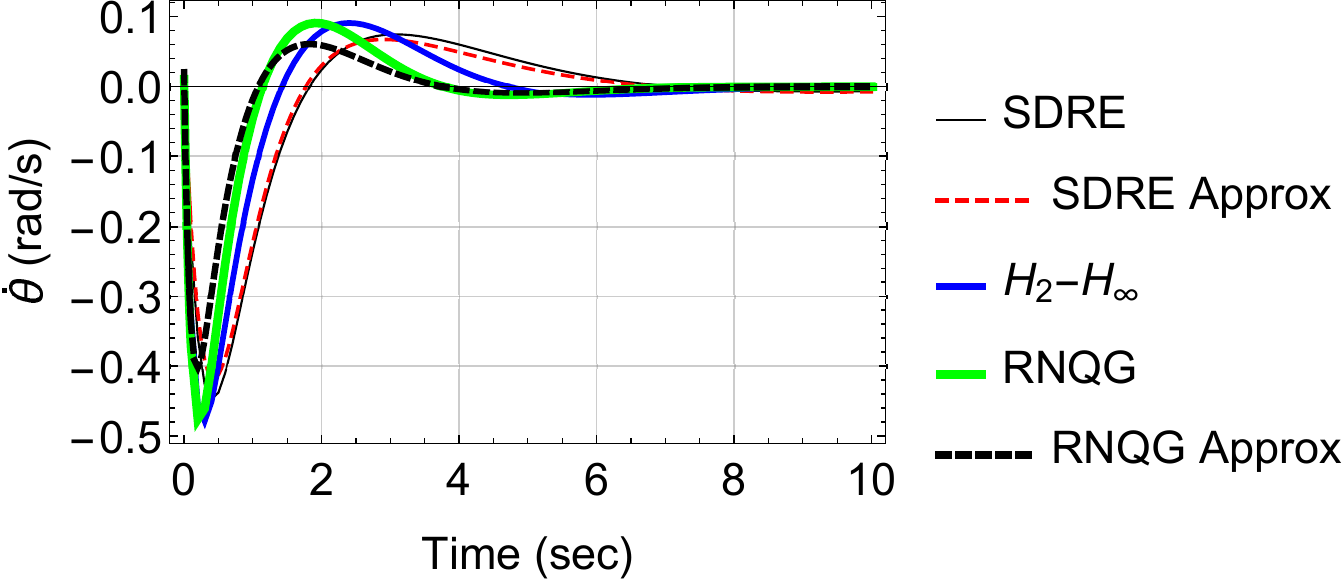} }}%
	\qquad
	\subfloat[The flywheel angular velocity.]{{\label{Simphi1}}{\includegraphics[width=8cm]{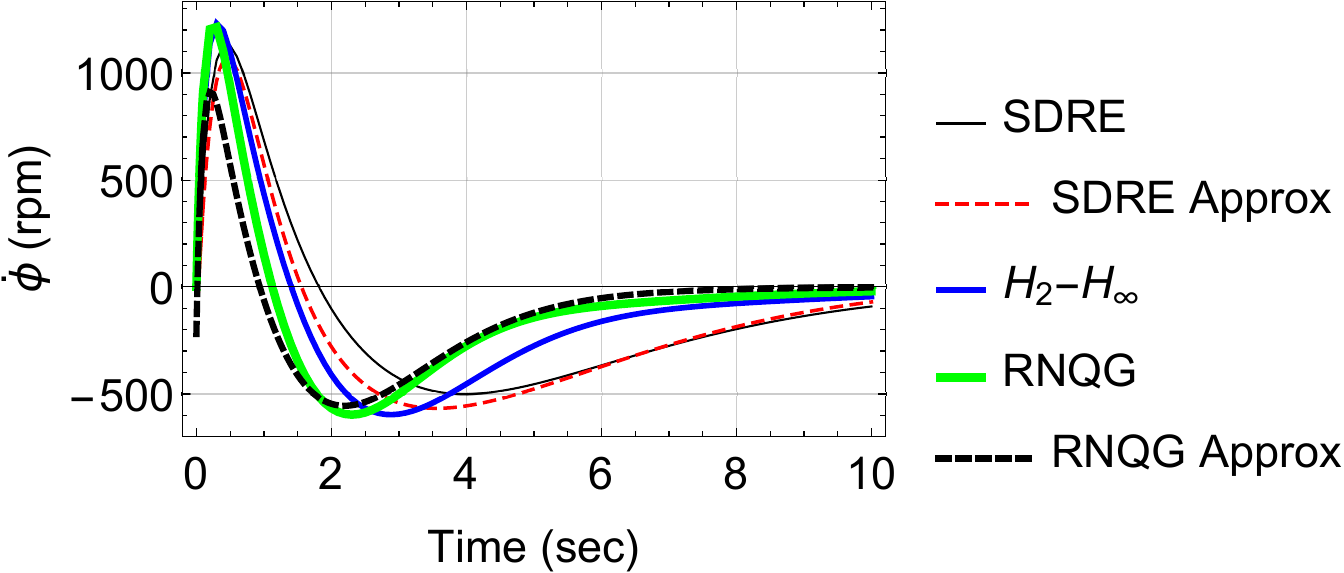} }}%
	\qquad
	\subfloat[The applied input.]{{\label{Simu1}}{\includegraphics[width=8cm]{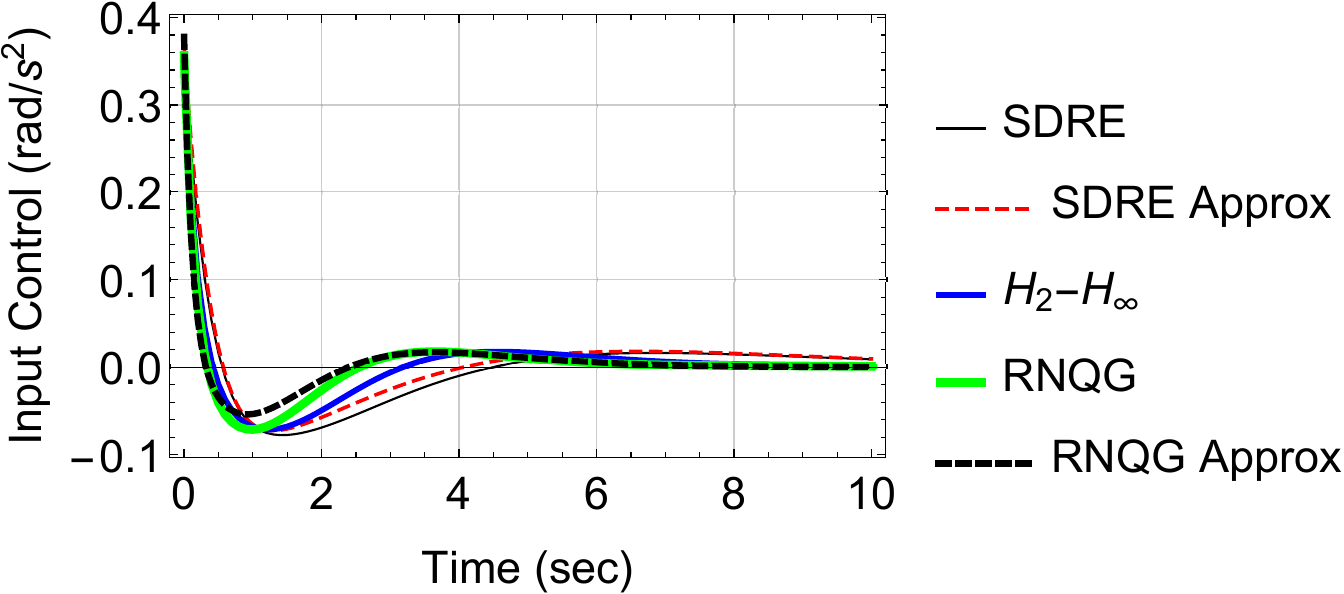} }}%
	\caption{Simulation results of system performances in the ideal condition.}
	\label{Simulation1}
\end{center}
\end{figure}

\begin{figure}[h!]
	\begin{center}
		\subfloat[The pendulum angle.]{{\label{Simtheta2}}{\includegraphics[width=8cm]{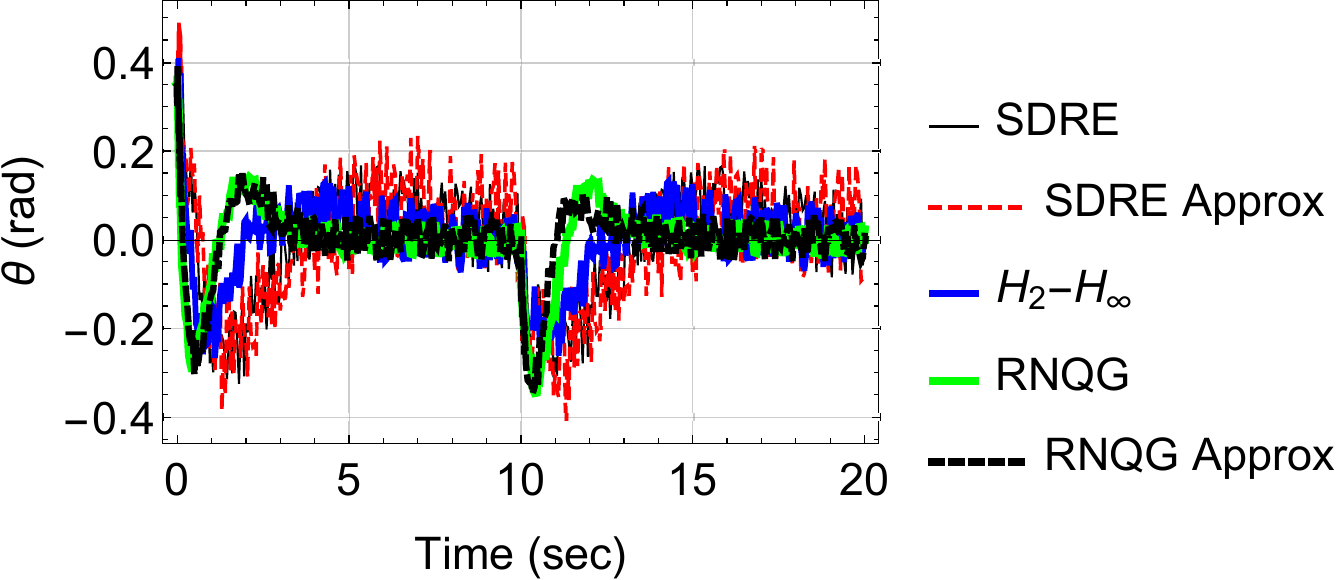} }}%
		\qquad
		\subfloat[The pendulum angle rate.]{{\label{Simdtheta2}}{\includegraphics[width=8cm]{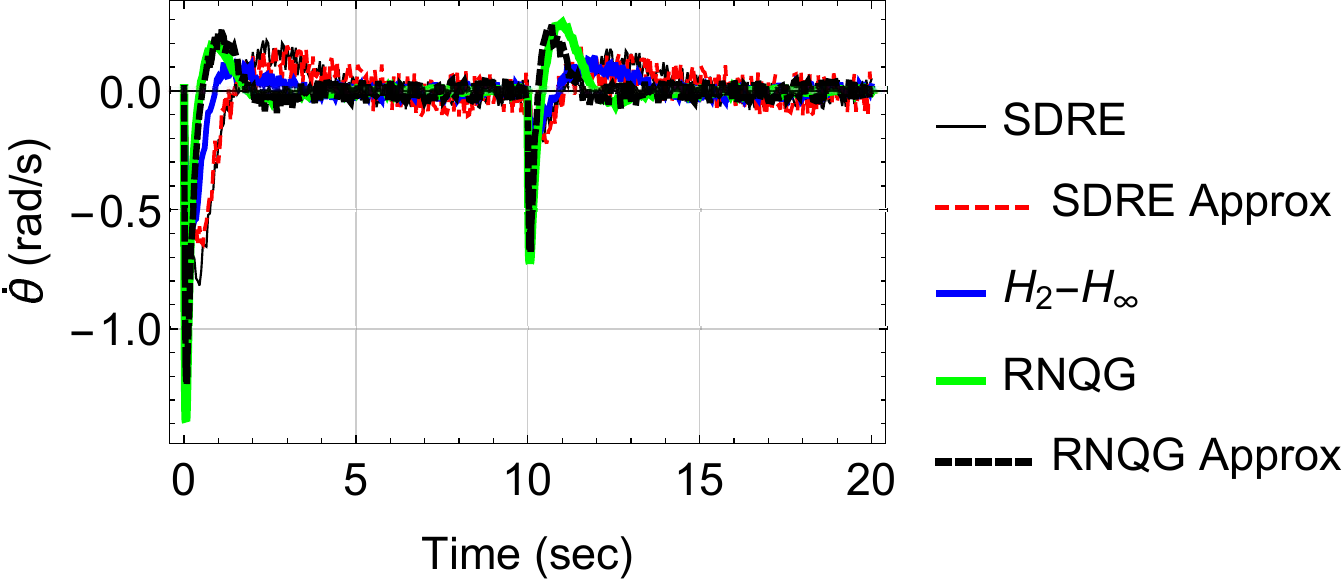} }}%
		\qquad
		\subfloat[The flywheel angular velocity.]{{\label{Simphi2}}{\includegraphics[width=8cm]{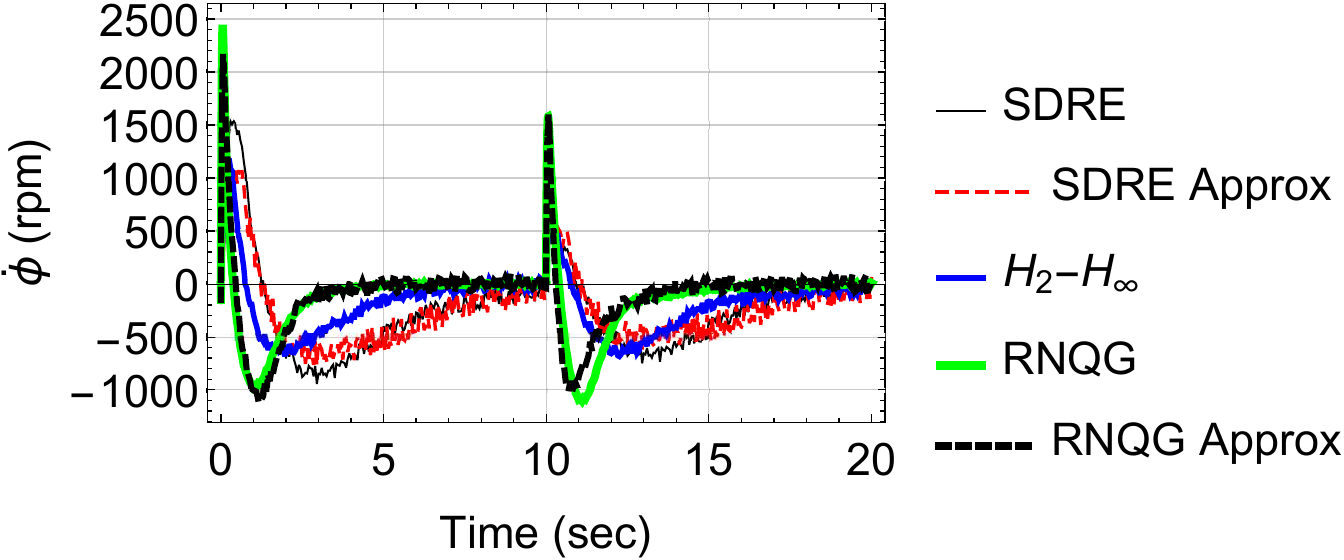} }}%
		\qquad
		\subfloat[The applied input.]{{\label{Simu2}}{\includegraphics[width=8cm]{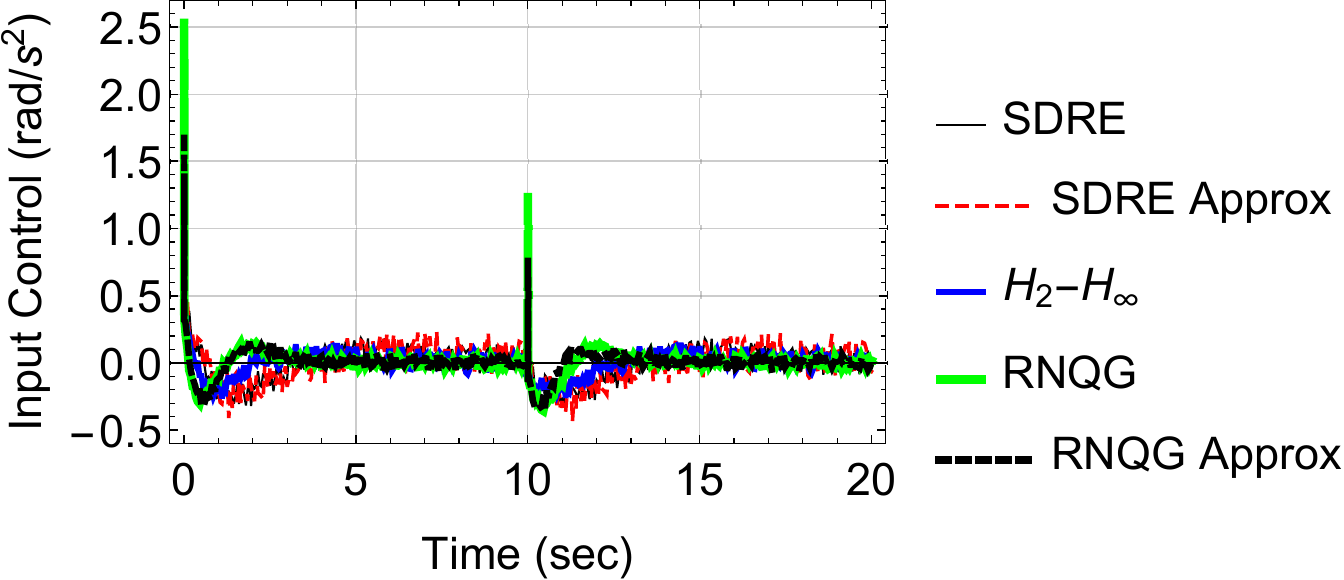} }}%
		\caption{Simulation results of system performances in the presence of disturbance and noises.}
		\label{Simulation2}
	\end{center}
\end{figure}

\begin{figure}[h!]
	\begin{center}
		\subfloat[The pendulum angle.]{{\label{Simtheta3}}{\includegraphics[width=8cm]{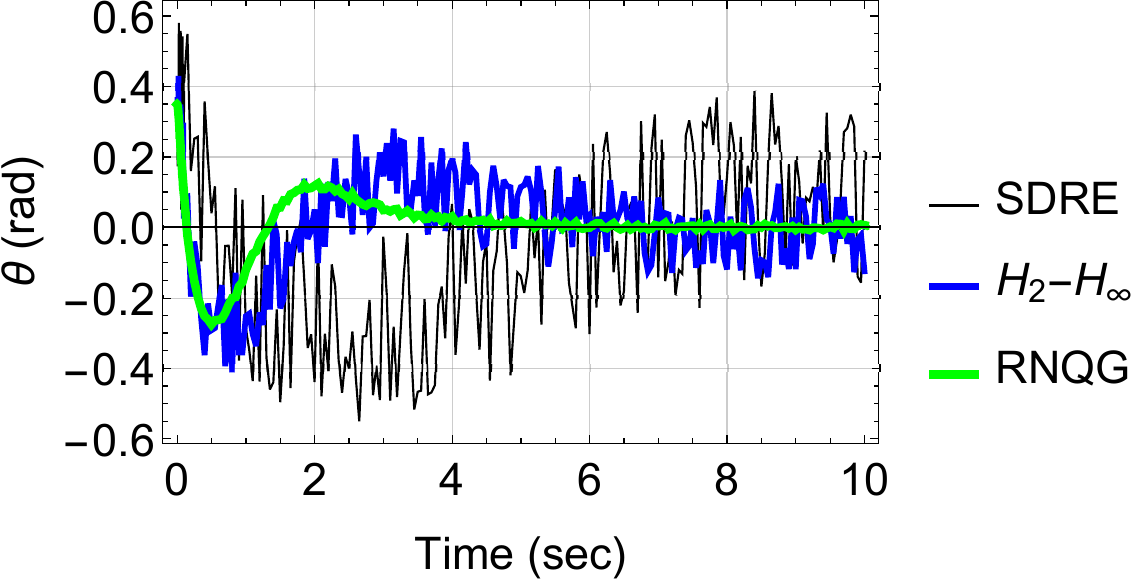} }}%
		\qquad
		\subfloat[The pendulum angle rate.]{{\label{Simdtheta3}}{\includegraphics[width=8cm]{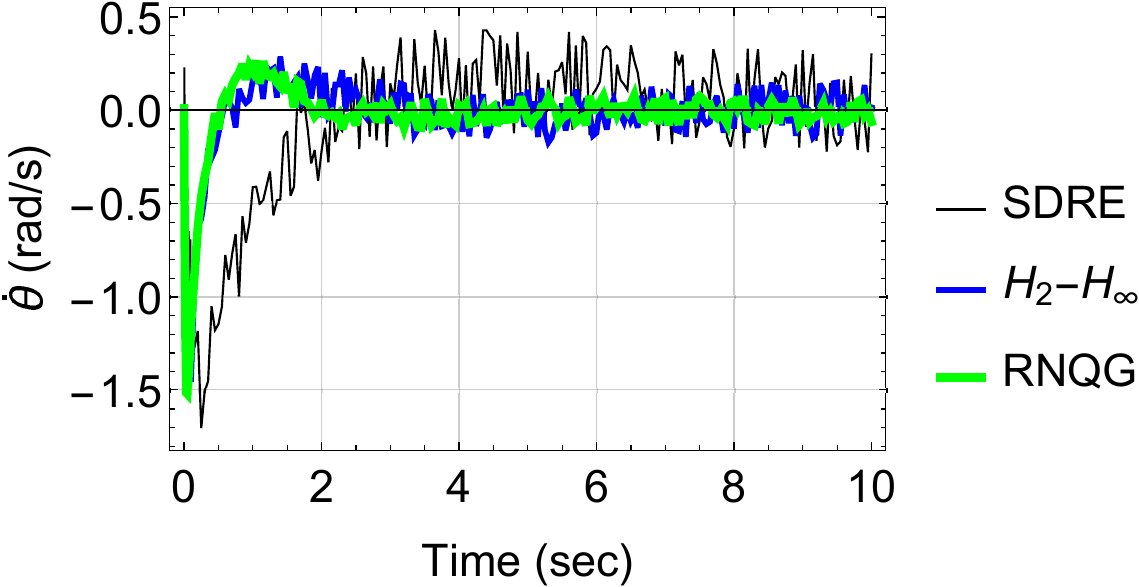} }}%
		\qquad
		\subfloat[The flywheel angular velocity.]{{\label{Simphi3}}{\includegraphics[width=8cm]{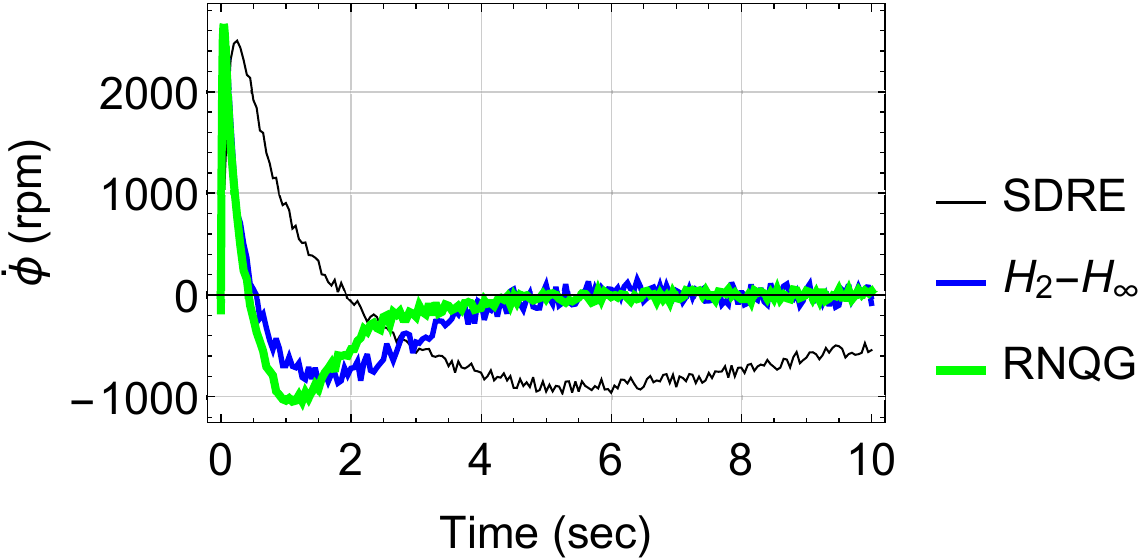} }}%
		\qquad
		\subfloat[The applied input.]{{\label{Simu3}}{\includegraphics[width=8cm]{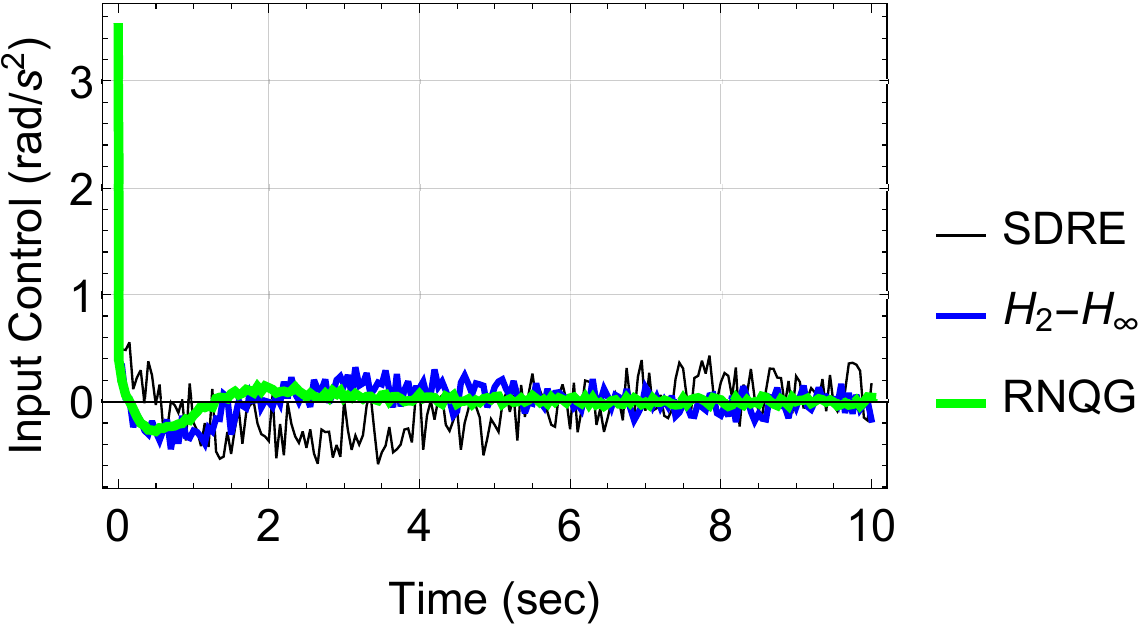} }}%
		\caption{Simulation results of system performances in the presence noises with magnitude 0.4 white noise .}
		\label{Simulation3}
	\end{center}
\end{figure}

To test the performance and illustrate the efficiency of the proposed controllers, two simulation examples are carried out using Wolfram Mathematica. First, a comparative simulation among controllers, the SDRE, SDRE approximation, $H_2-H_{\infty}$, RNQG, and RNQG approximation methods is presented where the system is considered as an ideal one. In the second example, the performance of the controllers are examined in the presence of disturbance and noise. The system parameters used in the simulations are presented in Table~\ref{details}. The comparisons between controllers are shown in Figs.~\ref{Simulation1}-\ref{Simulation3}. Also, we should note that the sampling time of the simulation is $\Delta t=0.01$ s. We used a controller sampling time of $\Delta t=0.01$ s during our simulations. 

\begin{table}[h]
	\centering
	\caption{The details of the system's parameters. }
	\label{details}
	\begin{tabular}{|c|c||c|c|}
		\hline
		parameter& value     & parameter   & value                         \\ \hline \hline
		$M_p$    & 0.6 $Kg$  &$L_e$          & 14 $cm$                       \\ \hline
		$M_w$    & 0.31 $Kg$ &$I_p$        & 0.0023 $Kg/m^3$               \\ \hline
		$L_G$    & 10 $cm$   &$I_w$        & 0.001 $Kg/m^3$                \\ \hline
	\end{tabular}
\end{table}

\subsection{Ideal System}
In the first simulation (Case 1), we consider an ideal system (\ref{EQMotion1}) and (\ref{EQMotion2}) with the following parameter matrices and initial conditions:

\begin{flalign}
& Q=\begin{pmatrix}
1+\theta^2 &0  &0  &0 \\ 
0 & 1+\phi^2 &0  &0 \\ 
0 & 0 & 1+\dot{\theta}^2 & 0\\ 
0 & 0 & 0 & 1+\dot{\phi}^2
\end{pmatrix}\,; \:\:\:\: R= I_1 ;  \nonumber\\
& S=I_4; \:\:\:\: \bar{x}_0=\{20^{\circ}, 0^{\circ}, 0.01\: \text{rad/s}, 0\: \text{rad/s}\}
\end{flalign} 
As can be seen, all controllers are stable and they converge to the desired values. It can be observed that the pendulum stabilizes in the vertically upright position quickly and smoothly after a minor overshoot. In addition, from the analysis of the simulation results, the responses of the control schemes and approximation method are similar as we expected.  Note that the tracking error can be remarkably reduced by selecting larger magnitudes for the components of the two matrices $Q$ and $R$ in all control schemes. This also leads to faster convergence speed. In addition, as one can see, the performance of the proposed RNQG is much better than that other controllers. 

\subsection{System subjected to disturbance and noise}
An external disturbance and zero mean white noise are introduced in these simulation runs. Where the external disturbance is applied after 10 seconds, and the noise is added to states as follows
\begin{flalign}
\bar{x} = \bar{x} + v
\end{flalign} 
where 
\begin{flalign}
v \sim N(0,q^2)
\end{flalign} 
We have selected two values 0.04 (Case 2) and 0.4 (Case 3) unit for q. The difference between RNQG and other controllers is more noticeable in this case (see Figs.~{\ref{Simulation2}} and {\ref{Simulation3}}).
As one can see, the RNQG outperforms the other controllers in presence of noise. The SDRE controller regulates $\dot{\phi}$ smoothly and does not exhibit significant overshoot unlike the RNQG scheme. When, we increased the magnitude of the noise, the performance of the proposed controller was shown better results such as convergence speed and eliminating the noise (see Fig.~{\ref{Simulation3}}). In this case, we only compared the SDRE, $H_2-H_{\infty}$, and RNQG to show the difference in their performances. 

\begin{table}[h]
	\centering
	\caption{Performance measurements of the proposed control schemes.}
	\label{comp1}
	\begin{tabular}{|c|c|c|c|c|}
		\hline
		& Controller       & IAE & ITAE & CEF \\ \hline
		\multirow{5}{*}{Case 1} & SDRE             &  123.03    &   333.88    &   0.041  \\ \cline{2-5} 
		& SDRE Approx.     &  116.99    &   305.83    &   0.038  \\ \cline{2-5}
		& $H_2-H_{\infty}$ &  112.01    &   227.27    &   0.035  \\ \cline{2-5} 
		& RNQG             &  110.21    &   164.08    &   0.035  \\ \cline{2-5}
		& RNQG Approx.     &  108.73    &   158.82    &   0.033  \\ \hline
		\multirow{5}{*}{Case 2} & SDRE             &  544.94    &   2673.56   &   0.088  \\ \cline{2-5}
		& SDRE Approx.     &  536.51    &   2655.02   &   0.084  \\ \cline{2-5} 
		& $H_2-H_{\infty}$ &  501.48    &   2320.34   &   0.054  \\ \cline{2-5} 
		& RNQG             &  411.47    &   1964.52   &   0.051  \\ \cline{2-5}
		& RNQG Approx.     &  407.83    &   1899.18   &   0.049  \\ \hline
		\multirow{3}{*}{Case 3} & SDRE             &  252.5    &   299.44   &   0.094  \\ \cline{2-5}
		& $H_2-H_{\infty}$ &  237.69    &   140.59   &   0.081  \\ \cline{2-5} 
		& RNQG             &  161.97    &   93.13   &   0.066  \\ \hline
		\end{tabular}
\end{table}

For more accurate and quantitative comparison of the proposed controller performances, three performance indicators are considered. These indicator measurements are: Integral Absolute Error (IAE), Integral Time Absolute Error (ITAE), and Control Energy Factor (CEF). They indicate the tracking error performance and the amount of the control effort of the system. These measurements can be calculated as follows: 
\begin{flalign}
& IAE=\sum_{i=1}^{3} \int_{0}^{t_{sim}} \left | q_i-q_i^d \right | dt \\
& ITAE= \sum_{i=1}^{3}\int_{0}^{t_{sim}} t\left | q_i-q_i^d \right | dt \\
& CEF= \int_{0}^{t_{sim}} u^2 dt \\
& \{q_1,q_2,q_3\}=\{\theta, \dot{\theta}, \dot{\phi}\} \nonumber
\end{flalign} 
where $t_{sim}$ is the total simulation time, $q$ is the state, and $q^d$ is the desired value for each state. Table~\ref{comp1} presents the performance measurements of the proposed controllers on the system for three cases.  In the ideal system, the outcome results showed similar behavior and had just 10 percent improvement. On the contrary, in the presence of noise, the improvement in tracking error became approximately 36 percent.

\section{Conclusions}
A novel optimal robust nonlinear controller has been investigated in this study. This control scheme was based on the SDRE approach. We showed that the generalized Riccati equation can be cast in the traditional form and solved with standard solvers. Then, an approximation method was presented, which was based on least squares technique and a Neural Network weight updated scheme. The objective of this approximation method was to facilitate real time implementation of the proposed methods. 

A flywheel-based inverted pendulum system was used to validate the control schemes. We also numerically compared the performances of the conventional SDRE, approximation approach, and $H_2-H_{\infty}$ controller with the proposed controller.  In the simulation, all controllers were able to stabilize the system. However, the performance of the RNQG was superior to the others. The proposed controller outperformed the others because of its capability in dealing with disturbance and process noise.

\bibliographystyle{plain}        
\bibliography{paper}           
 
\end{document}